\documentstyle[12pt]{article}

\newcommand{\be}{\begin{equation}}
\newcommand{\ee}{\end{equation}}
\begin{document}

\topmargin 0pt
\oddsidemargin=-0.4truecm
\evensidemargin=-0.4truecm

\renewcommand{\thefootnote}{\fnsymbol{footnote}}

\vspace*{-2.0cm}

\vspace*{0.5cm}
\begin{center}
{\Large \bf Induced charges as probes\\ 
of low energy effective theories}

\vspace{0.5cm}

{\large
V. A. Rubakov
\footnote{E-mail: rubakov@ms2.inr.ac.ru}}

\vspace*{0.4cm}

{\em Institute for Nuclear Research of the Russian Academy of Sciences,\\ 
60th October Anniversary Prospect, 7a, Moscow 117312, Russia
}
\end{center}

\vglue 0.8truecm
\begin{abstract}
We suggest that the correspondence between gauge theories
strongly coupled in the infrared and 
their low energy effective theories
may be probed by introducing topologically non-trivial 
background scalar fields. We argue that 
 one loop expressions for
the global charges induced in vacuum
by these background fields are  in some cases exact 
in the fundamental theory, and hence should be matched in
the effective theory.
These matching conditions  are sometimes
inequivalent to 't Hooft ones. A few examples of induced 
charge matching are presented.
\end{abstract}
\vspace{1.cm}

\renewcommand{\thefootnote}{\arabic{footnote}}
\setcounter{footnote}{0}

1. Low energy effective theories  are used to describe low
energy physics inherent in fundamental, 
``microscopic'' theories which are strongly
coupled in the infrared. An important guide to infer the
effective theories is provided by the 't Hooft anomaly matching
conditions \cite{tHooft} (and their discrete analogs \cite{discrete}).
One way to understand the anomaly matching  is to introduce
background gauge fields corresponding to (a subgroup of) the flavor
symmetry group; the anomalies in the flavor currents are then
proportional to the topologicalcal charge densities of the background
gauge fields.  This topological property is closely related to the
Adler--Bardeen theorem that guarantees that the anomalies are equal in
the original, ``microscopic'' theory and its low energy partner.

In this paper we suggest that the low energy theories may also be
probed by introducing slowly varying scalar, rather than vector,
background fields with topological properties. The global charges
induced in vacuum by these fields \cite{GW} are often (though not always)
unambiguously calculable at one loop 
in the fundamental theory, and are
proportional to the topological charges of the background. This
topological property suggests, by analogy to the triangle anomalies,
that the non-renormalization theorem similar to that of Adler and 
Bardeen
should in many cases
hold for induced charges, and that in these cases
the induced charges, calculated at one loop 
in low energy and fundamental theories,  should
match.

Though induced charges and triangle anomalies are closely related to
each other, they probe somewhat different aspects of the
correspondence between fundamental and low energy theories. The
background scalar fields provide 
masses to (some of) the fermions of the
fundamental theory, so the induced charges probe the respond of the
low energy theory to these masses. 
In some cases --- in particular, in 
supersymmetric gauge theories --- this respond is
well understood, so the induced charge matching adds nothing 
to the analysis of the low energy theories. Indeed,
we will see that there are very simple
sufficient conditions that insure induced charge matching provided
the 't Hooft matching holds; these 
sufficient conditions are automatically
satisfied in supersymmetric gauge theories. On the other hand,
we will see that 
in non-supersymmetric models including those
in which
supersymmetry is slightly broken by  small soft masses,
induced charge matching places 
constraints on the low energy theories that are
not equivalent to the 't Hooft constraints.

In this paper we first 
present arguments supporting our conjecture of the absence of quantum 
corrections to induced charges, taking non-supersymmetric QCD as an 
example. We then
give a few other examples of the induced charge
matching and see that it sometimes occurs in a fairly non-trivial way.
We conclude by pointing out that induced charges
may suffer  infrared problems in some theories, 
so the induced charge matching
between low energy and fundamental theories may  not occur.

2. We begin with conventional $SU(N_c)$ QCD
with $N_f$ massless flavors. Let us introduce background fields
$m^{\tilde p}_{q}({\bf x})$ which are time independent and slowly
vary in space. These are $N_0 \times N_0$ matrices;
hereafter the indices $p,q,r; \tilde{p}, \tilde{q}, \tilde{r}$ run
from $1$ to $N_0$ with $N_0 \leq N_f$. Let these fields couple to $N_0$
flavors of quarks and anti-quarks in the following way,
\[
      m^{\tilde p}_{q}({\bf x})\tilde{\psi}_{\tilde{p}}  \psi^q \;
      + \; \mbox{h.c.}
\]
where $\psi^i$ and $ \tilde{\psi}_{\tilde{j}}$ are left-handed 
quark and anti-quark fields, respectively. We assume 
for definiteness that the background fields have the following form
\begin{equation}
   m^{\tilde p}_{q}({\bf x}) = m_0 U^{\tilde p}_{q}({\bf x})
\label{22*}
\end{equation}
where $m_0$ is a constant and $U$ is an $SU(N_0)$ matrix at each point 
${\bf x}$. We restrict the form of the
background even further
by requiring that $U({\bf x})$ is independent of coordinates
 at spatial infinity;
by a global $SU(N_f)_L$ rotation
\[
     U(\bf x) \to 1 \;\; as \;\; |{\bf x}| \to \infty
\]
Under these conditions, the background fields are characterized by
the topological charge
\[
    N[U] = \frac{1}{24 \pi^2} \int~d^3 x~ \epsilon^{ijk}
           \mbox{Tr} \left( U\partial_i U^{\dagger}
                 U\partial_j U^{\dagger}  U\partial_k U^{\dagger}        
            \right)
\]
The background fields give ${\bf x}$-dependent
masses to $N_0$ quarks, while $(N_f - N_0)$
flavors remain massless. These fields also explicitly break the flavor
group down to  \\
$SU(N_f - N_0)\times SU(N_f - N_0) \times U(1)_B \times U(1)_8^f$
where $U(1)_B$ is the baryon number (we assign baryon number 1 to each
quark) and $U(1)_8^f$ is a vectorial subgroup of the original
$SU(N_f)\times SU(N_f)$
flavor group, whose (unnormalized) generator  is
\[
    T_8^f = \mbox{diag}\left( 1, \dots, 1, -\frac{N_0}{N_f - N_0},\dots,
                  -\frac{N_0}{N_f - N_0}\right)
\]
( $U(1)_8^f$ is absent if $N_0 = N_f$).

The background fields $m({\bf x})$ induce global charges in vacuum.
We are interested in the global symmetries which are unbroken by the
background fields and under which the massive quarks transform
non-trivially. These are the baryon number and $T_8^f$. The one-loop
calculation \cite{GW} gives for slowly varying $m({\bf x})$
\begin{equation}
    \langle B \rangle = N_c N[U]
\label{5*}
\end{equation}  
\begin{equation}
 \langle T_8^f\rangle = N_c N[U]
\label{5**}
\end{equation}
As the right hand side of these relations equals the topological number,
up to a color factor, we suggest that eqs. (\ref{5*}) and (\ref{5**})
are valid in full quantum theory.

To substantiate this conjecture,  
let us discuss the relation between induced charges and 
triangle anomalies; we consider induced baryon number as an example.
 The ${\bf x}$-dependence of the background field 
$m({\bf x})$ can be removed at the expence of modification of the
gradient term in the quark Lagrangian.
 Namely, after the $SU(N_0)_L$
rotation of the left-handed quark fields $\psi^p$, 
$\psi (x) \to U^{-1}({\bf x}) \psi (x)$,
$\tilde{\psi}(x) \to \tilde{\psi}(x)$, first $N_0$ quarks and anti-quarks 
have ${\bf x}$-independent masses $m_0$, and the gradient term
of these quarks becomes 
$\bar{\psi}~ i \gamma \cdot\left(D + \frac{1- \gamma^5}{2} A^L\right)
\psi$  where
$A_0^L =0$, $A^L_i = U \partial_iU^{-1}$, $D_{\mu} $ contains
dynamical gluon fields, and we temporarily
switched to four-component notations.
This addition to the gradient term may be viewed as
the interaction of massive quarks with the background pure gauge 
vector fields corresponding to $SU(N_0)_L$ subgroup of the flavor group;
these background fields are small and slowly vary in space.

Now, consider an adiabatic process (either in Minkowskian
or in Euclidean space-time) in which the background vector
fields $A^L (x)$  (in the gauge $A_0 = 0$) change 
in time from $ A^L_i = 0$
to $A^L_i = U \partial_iU^{-1}$ always varying slowly 
in space and vanishing at spatial infinity (an example of such a 
process is an instanton of large size).
 Suppose that this process 
begins with the system in the ground state which has zero induced 
charges because of the triviality of the background.  As the 
background vector fields interact with massive degrees of freedom only,
the system remains in its ground state in the entire process, at least
order by order in perturbation theory. The induced baryon number in the
final state --- the quantity we are interested in --- is equal to
$<B> = \int~d^4 x~ \partial_{\mu} j^B_{\mu}$ which in turn is
determined by the anomaly in the gauge-invariant baryonic current 
$j^B_{\mu}$. Hence, we recover eq. (\ref{5*}):
\begin{equation}
<B> = \frac{N_c}{16 \pi^2} \int~d^4 x~ F^L_{\mu\nu}
         \tilde{F}^L_{\mu\nu} 
       = N_c N[U]
\label{11*}
\end{equation}
This observation relates induced charges and anomalies and
strongly suggests that induced charges do not receive radiative
corrections in the fundamental, ``microscopic'' theory.

This argument is still basically perturbative. One may wonder whether
non-perturbative effects such as fermion level crossing might
make the final state 
of the adiabatic process different from the ground state, i.e.,
whether the final state might actually contain excitations carrying
non-zero net baryon number. In that case the baryon number induced in the
ground state by the background field 
$A^L_i ({\bf x}) = U \partial_i U^{-1} ({\bf x})$ would be different 
from eq. (\ref{11*}), as the anomaly determines the {\it total}
change in the baryon number. To argue that this does not happen, we
note that the appearance, in the final state, of excitations with
non-zero net baryon number would show up as a non-vanishing index
of the four-dimensional
Euclidean Dirac operator ${\cal D}[A^L] = \gamma 
\cdot \left(D + \frac{1-\gamma^5}{2} A^L (x) \right) + m_0$, so
that the vacuum-to-vacuum amplitude would vanish while matrix elements
of baryon number violating operators between the 
initial and final vacua
would not. However, for arbitrary gluon fields, the 
eigenvalues $\omega$ of the operator ${\cal D}[A^L = 0] = \gamma 
\cdot D + m_0$ obey $|\omega| > m_0 $ (the Euclidean operator
$\gamma \cdot D $ is anti-Hermitean) so the operator ${\cal D}[A^L]$
has no zero modes when the background fields $A^L (x)$ are small
($A^L (x) \ll m_0$ at all $x$) and slowly vary in space-time.
This argument implies that eq. (\ref{11*}) is valid in full quantum 
theory even at $m_0 < \Lambda_{QCD}$. Although the situation in
theories with colored scalars is more complicated, it is likely 
that analogous arguments may be designed in those theories as well.

Let us see that the low energy effective theory of QCD --- the
non-linear sigma model --- indeed
reproduces eqs. (\ref{5*}) and (\ref{5**}).
In the absence of the background fields, the non-linear sigma model
action contains only derivative terms for the $SU(N_f)\times SU(N_f)$
matrix valued sigma-model field $V(x)$, 
including the usual kinetic term and
the Wess--Zumino term. The background field $m({\bf x})$ 
introduces a potential
term into the low energy effective Lagrangian,
\[
           \Delta L_{eff} =
         \mbox{Tr}\left( m^{\dagger}V + V^{\dagger}m \right)
\]
For slowly varying $m$, 
the effective potential is  minimized at
\begin{equation}
 V({\bf x}) =
\left(
\begin{array}{cc}
   U({\bf x})  & 0 \\
   0 &  {\bf 1}
\end{array}
\right)   
\label{6*}
\end{equation}
Hence, the induced baryonic charge appears at the classical level
\cite{polychron}; as the baryonic charge of $V({\bf x})$ is equal to
its topological number $N[V]$ times $N_c$, the induced baryonic charge is
indeed given by eq.(\ref{5*}). Likewise, it follows from the structure of
the Wess--Zumino term that the $T^f_8$ current of the configuration
of the form (\ref{6*}) is (cf. \cite{farhi})
\[
    j_{8,\mu}^f = \frac{N_c}{24 \pi^2}  \epsilon_{\mu\nu\lambda\rho}
           \mbox{Tr} \left( U\partial_{\nu} U^{\dagger}
                 U\partial_{\lambda} U^{\dagger}  U\partial_{\rho}
            U^{\dagger}        
            \right)
\]
so the  $T^f_8$ charge of the configuration (\ref{6*})
is given by eq.(\ref{5**}).

We see that the induced charges in QCD and its low energy effective theory
match  rather trivially. The way the induced charges match becomes more 
interesting when low energy theories contain massless fermions.

3. Let us now consider supersymmetric QCD with $N_c$ colors and $N_f$
flavors. To be specific, we discuss the case $3N_c > N_f > N_c +3$.
This theory exhibits the Seiberg duality \cite{Seiberg-du}: the
fundamental theory contains the superfields of quarks $Q^i$, anti-quarks
$\tilde{Q}_{\tilde{j}}$ and gluons, while its effective low energy
counterpart at the origin of moduli space is an $SU(N_f - N_c)$ 
magnetic gauge theory with magnetic quarks $q_i$, magnetic anti-quarks
$\tilde{q}^{\tilde{j}}$ and mesons $M^i_{\tilde{j}}$ with the 
superpotential $qM \tilde{q}$. 

Let us probe this theory by adding the scalar background fields
$m^{\tilde{q}}_p({\bf x})$ with the same properties as above, i.e.,
by introducing the term
\begin{equation} 
   m^{\tilde{q}}_p({\bf x}) \tilde{Q}_{\tilde{q}} Q^p    
\label{31*} 
\end{equation}
into the superpotential of the fundamental theory. Let us take for 
definiteness\\ 
$2 \leq N_0 < N_f - N_c - 1$. The calculation of the induced 
baryon and $T_8^f$ charges in the fundamental theory proceeds as above,
and we again obtain eqs.(\ref{5*}) and (\ref{5**}).

Let us turn now to the effective low energy theory. For slowly varying
$m({\bf x})$, the term (\ref{31*}) translates into $\mbox{Tr}(mM)$,
so the total superpotential of the magnetic theory is
\begin{equation} 
   qM\tilde{q} + \mu_0\mbox{Tr}(mM)
\label{32+} 
\end{equation} 
where $\mu_0$ is the dimensionfull parameter inherent in the magnetic 
theory.
The ground state near the origin of the moduli 
space has the following non-vanishing ${\bf x}$-dependent
expectation values\footnote{Hereafter we use the same notations
for superfields and their scalar components.} of the magnetic 
quarks and anti-quarks,
\begin{equation} 
     < q^p_q> \;= \mu_q^p\;, \;\;p= 1, \dots, N_0, \;\; q=1, \dots, N_0
\label{32*}
\end{equation} 
(here the upper and lower indices refer to magnetic color and
flavor, respectively) 
\begin{equation} 
    <\tilde{q}^{\tilde{q}}_p> \; =
   \tilde{\mu}^{\tilde q}_p\;, \;\;p= 1, \dots, N_0, \;\; 
     \tilde{q}=1, \dots, N_0
\label{32**} 
\end{equation}
(here the lower index refers to magnetic color). 
The expectation values obey
\[
    \tilde{\mu}^{\tilde r}_p ({\bf x}) \mu_q^p ({\bf x}) = 
        - \mu_0 m_q^{\tilde{r}} ({\bf x})
\]
They also satisfy the $D$-flatness condition at each point in space, 
$         \mu^{\dagger q}_p\mu_q^r  =
 \tilde{\mu}^{\tilde q}_p \tilde{\mu}^{\dagger r}_{\tilde q}$.
With our choice of background fields, eq.(\ref{22*}), one has
\[
     \mu = \pm \sqrt{\mu_0 m_0}~W({\bf x})\; , \;\;
     \tilde{\mu} = \mp \sqrt{\mu_0 m_0}~\tilde{W}({\bf x})
\]
where $W$ and $\tilde{W}$ are $N_0 \times N_0$ unitary 
matrices\footnote{At $m = m_0\cdot {\bf 1}$, the matrices $\mu$ and 
$\tilde{\mu}$ are proportional to  $N_0 \times N_0$ unit matrix, up 
to magnetic color rotation. At $m = m_0 U({\bf x})$ one has
$\mu = \pm \sqrt{\mu_0 m_0}~U_c U$, 
$\tilde{\mu} = \mp \sqrt{\mu_0 m_0}~U_c^{\dagger}$
where $U_c({\bf x})$ is a slowly varying matrix belonging to
$SU(N_0)$ subgroup of the magnetic color group. The explicit form of
$U_c({\bf x})$ is to be found from the minimization of the gradient 
energy, and it is not important for our purposes.} obeying
\begin{equation} 
      \tilde{W}W({\bf x}) = U({\bf x})
\label{23++} 
\end{equation} 
Since the gradient energy has to vanish at spatial 
infinity, $W({\bf x})$ and  $\tilde{W}({\bf x})$ are constant at
$|{\bf x}| \to \infty$, so they can be characterized by their
winding numbers $N[W]$ and  $N[\tilde{W}]$. Because of eq. (\ref{23++})
one has
\[
      N[W] + N[\tilde{W}] = N[U]
\]
In this ground state, the magnetic color is broken down to
$SU(N_f - N_c - N_0)$. At small $m_0$, the ground state
(\ref{32*}), (\ref{32**}) is close to the origin of the moduli
space, so the magnetic description is  reliable.

Both the baryon number and $T_8^f$ are broken in this vacuum.
However, there exist combinations of these generators and
magnetic color generators that remain unbroken.
Recalling \cite{Seiberg-du} that the baryon number of magnetic quarks
equals $N_c/(N_f - N_c)$ and that the magnetic quarks and anti-quarks
transform as $(\bar{N}_f, 1)$ and $(1,N_f)$, respectively,
under the global $SU(N_f)\times SU(N_f)$ group, 
the unbroken generators are
\begin{equation} 
     B' = B - \frac{N_c}{N_f - N_c} T_8^{mc}
\label{33*}
\end{equation}
 \begin{equation} 
     T'_8 = T_8^f + T_8^{mc}
\label{33**}
\end{equation}
where $ T_8^{mc}$ is the following generator of the magnetic color
\[
    T_8^{mc} = \mbox{diag} \left(1,\dots,1, -\frac{N_0}{N_f - N_c - N_0},
    \dots -\frac{N_0}{N_f - N_c - N_0} \right)
\]
As the fundamental quarks and gluons are singlets under magnetic color,
the induced charges $<B'>$ and $<T'_8>$ calculated in the magnetic theory
should match eqs. (\ref{5*}) and (\ref{5**}). Let us check that this is
indeed the case\footnote{This example explains why we prefer
to deal with the induced {\it charges} rather than the induced 
{\it currents}. The currents corresponding to the generators (\ref{33*}),
(\ref{33**}) may be quite complicated in the fundamental theory, so 
the calculation of their expecation values --- induced currents ---
does not seem possible. On the other hand, mapping of charges in the
fundamental and effective theories is dictated by symmetries alone.}.

The induced charges appear in the magnetic theory through 
${\bf x}$-dependent 
mass terms of fermions. These are generated by the expectation values 
(\ref{32*}), (\ref{32**}). The mass terms coming from the superpotential
(\ref{32+}) are
\begin{equation} 
    \tilde{\mu}^{\tilde{p}}_q ({\bf x})
\Psi_{\tilde{p}}^i \psi_i^q
   + \mu^{p}_q ({\bf x})
\tilde{\psi}^{\tilde{j}}_p \Psi_{\tilde{j}}^q
\label{44**} 
\end{equation}
where $\Psi$, $\psi$ and $\tilde{\psi}$ are fermionic components
of mesons, magnetic quarks and magnetic anti-quarks, respectively.
The gauge interactions give rise to other mass terms,
\begin{equation} 
   \mu^{\dagger p}_q ({\bf x})
\psi^a_p \lambda^q_a
   - \tilde{\mu}^{\dagger p}_{\tilde{q}}  ({\bf x})
\lambda^a_p 
      \tilde{\psi}_a^{\tilde q} 
\label{44*} 
\end{equation}
where $\lambda^b_a$ is the gluino field, $a,b = 1,\dots, (N_f - N_c)$
are magnetic color indices.

To calculate the induced baryon number $<B'>$ we observe that the only 
fermions carrying non-zero $B'$ are magnetic quarks $\psi_i^{\alpha}$
with $i = 1,\dots, N_f$, \\
$\alpha = (N_f - N_c - N_0 + 1), \dots, (N_f - N_c)$,
magnetic anti-quarks $\tilde{\psi}^{\tilde{j}}_{\alpha}$ and
gluinos $\lambda^{\alpha}_p$, $\lambda_{\alpha}^p$. Their $B'$-charges
are
\[
   \psi_i^{\alpha}: \; \frac{N_c}{N_f - N_c} -
      \frac{N_c}{N_f - N_c}\left(-\frac{N_0}{N_f - N_c - N_0}  
      \right) = \frac{N_c}{N_f - N_c - N_0}
\]
\[
   \lambda^{\alpha}_p : \; \frac{N_c}{N_f - N_c - N_0}
\]
\[
  \tilde{\psi}^{\tilde{j}}_{\alpha}\;, \;\; \lambda_{\alpha}^p
     : \; -\frac{N_c}{N_f - N_c - N_0}
\]
Hence, the induced $B'$ is due to the ${\bf x}$-dependent mass term
(\ref{44*}) and is equal to
\[
  <B'> =  -\frac{N_c}{N_f - N_c - N_0} \cdot (N_f - N_c - N_0)
         \left( N[W^{\dagger}] +  N[\tilde{W}^{\dagger}] \right)
\]
Due to eq. (\ref{23++}) it indeed coincides with $N_c N[U]$, the
induced
baryon number calculated in the fundamental theory.

The induced charge $<T'_8>$ is calculated in a similar way.
The relevant $T'_8$ charges of magnetic quarks are
\[
   \psi_p^{\alpha}: \; -\frac{N_f - N_c}{N_f - N_c - N_0}
\]
\[
\psi_p^{u}: \; \frac{N_f}{N_f - N_0}\;, \;\; u = (N_0 +1), \dots, N_f
\]
and similarly for magnetic anti-quarks, gluinos and mesons.
We find that both ${\bf x}$-dependent mass terms, (\ref{44**}) and
(\ref{44*}), contribute to $<T'_8>$, and obtain
\[
<T'_8> =  \frac{N_f}{N_f - N_0}\cdot (N_f - N_0)
         \left( N[W] + N[\tilde{W}] \right)
    +
       \frac{N_f - N_c}{N_f - N_c - N_0} \cdot (N_f - N_c - N_0)
        \left( N[W^{\dagger}] + N[\tilde{W}^{\dagger}] \right)
\]
This is equal to $N_c N[U]$, so the induced $T'_8$ charges also
match in the fundamental and low energy theories.

4. As our last example, let us consider supersymmetric
QCD with small soft masses of scalar quarks, $m_Q^2$, that explicitly
break supersymmetry \cite{Peskin}. We again probe this theory by
introducing the term (\ref{31*}) into the superpotential. The restrictions
on $N_f$, $N_c$ and $N_0$ are the same as in the previous example.

The induced charges, as calculated in the fundamental theory,
are still given by eqs. (\ref{5*}) and (\ref{5**}). The low energy 
theory near the origin is still the magnetic theory, but now
with soft mass terms of scalar mesons and scalar magnetic quarks
\cite{Peskin}. The scalar potential of the magnetic theory near the 
origin
at small $m_Q^2$ is determined both by the superpotential (\ref{32+}) 
and these soft terms,
\begin{equation} 
   V(M, q, \tilde{q}) = |\tilde{q}q + \mu_0 m|^2 + |qM|^2
                       + |M\tilde{q}|^2 + m_M^2 M^{\dagger} M
       + m_q^2 (q^{\dagger}q + \tilde{q}^{\dagger} \tilde{q})
       + D\mbox{-terms}
\label{52*} 
\end{equation}
where $m_M^2$ and $m_q^2$ are proportional to $m_Q^2$. Were the soft
terms in eq. (\ref{52*}) positive, the ground state of this theory at
$m_q^2 > \mu_0 m_0$ would be at the origin, 
$<q> = <\tilde{q}> = <M> =0$. The masses of fermions in the magnetic 
theory would vanish, the induced charges $<B>$ and $<T_8^f>$ would be
zero, so the induced charge matching would not occur. Hence, the induced
charge matching  {\it requires} 
that either $m_q^2$ and/or $m_M^2$ 
are negative, so that the ground state even at $m_0=0$
is  away from the origin,
or $m_q^2 = 0$,  $m_M^2 \geq 0$ with the ground state being the same 
as in the prevous example. This is in accord with explicit calculations:
it has been found in ref. \cite{Rattazzi} (see also ref. \cite{another})
that $m_q^2 < 0$ at $N_c + 1 < N_f < 3N_c/2$,
i.e., when the magnetic theory 
is weakly coupled,
while at $3N_c/2 \leq N_f < 3N_c$ one has $m_q^2 = m^2_M = 0$
\cite{conformal}. We conclude that the induced charge matching provides
qualitative understanding of these results\footnote{The same phenomenon
occurs in softly broken supersymmetric theories with $SO(N_c)$ and 
$Sp(2k)$ gauge groups and fundamental quarks at $N_f$, $N_c$
and $k$ such that the Seiberg duality holds \cite{DG}.}.

It is worth noting that there exists an example \cite{Rattazzi}
where soft masses of scalar quarks single out the 
vacuum at the origin of the moduli space (in the absence of the
background fields $m ({\bf x})$). This is the theory with
$Sp(2k)$ gauge group and $2k +4 = 2N_f$ quarks $Q_i$, 
$i=1,\dots, 2N_f$, in the fundamental
representation. The low energy effective theory \cite{sp2k}
contains antisymmetric mesons $M_{ij}$ and has superpotential
$\mbox{Pf}~M$. One can probe this theory by adding 
${\bf x}$-dependent mass terms 
$m^{p\tilde{q}}({\bf x}) Q_{\tilde{q}}Q_p$ where $p=1,\dots, N_f$, 
$\tilde{q} = (N_f+1),\dots, 2N_f$. In the theory without soft
supersymmetry breaking, the induced charges match much in the same way
as in the previous examle: scalar mesons obtain the expectation
values \\
$<M_{\tilde{q}p} ({\bf x})> \propto m^{\dagger}_{\tilde{q}p} ({\bf x})$
which give ${\bf x}$-dependent masses to fermionic mesons. After
the soft scalar quark masses are introduced, the scalar
potential of the low energy theory contains soft meson masses,
$m_M^2 M^{\dagger}M$ where 
$m_M^2 > 0$ at $k>1$ \cite{Rattazzi}. At first sight, this ruins the
induced charge matching at small $m_0$, as the ground state appears to 
be at $M=0$ and no ${\bf x}$-dependent masses of fermionic mesons seem
to be generated. However, the symmetries of the theory allow for a
linear supersymmetry breaking term in the scalar potential,
$m_Q^2 f(m_Q^2, m_0) mM$, which shifts the ground state to 
$<M> \propto m^{\dagger}$ and in this way restores induced charge 
matching\footnote{Note that a term linear in $q$ is not possible in
the scalar potential of the magnetic theory in the previous 
example, as the
magnetic quarks carry magnetic color.}. Hence, we argue that this
linear term is indeed generated in the low energy theory.

5. 
To conclude this paper, let us make two remarks. 
Let us come back
to the adiabatic process leading to eq. (\ref{11*}), and again discuss
induced baryon number as an example.
Our first remark is that the same
adiabatic process may be considered 
within the low energy effective theory.
The induced baryon number is now related to the anomaly in the
effective theory,  {\it provided}
all low energy degrees of freedom interacting with $SU(N_0)$
gauge fields become massive upon introducing the mass $m_0$
to $N_0$ flavors of fundamental quarks. As the 
$U(1)_B \times SU(N_0)_L \times SU(N_0)_L$ anomalies are the
same in the fundamental and low energy theories,
the induced baryon numbers match automatically in that 
case. Hence, a sufficient condition for induced charge matching is that
no low energy degrees of freedom transforming non-trivially under
a subgroup of the flavor group remain massless when this subgroup is
explicitly broken by masses of some fermions of the fundamental 
theory. This property is certainly valid in supersymmetric theories
where no phase transition is expected to occur as the masses of some of
the flavors are changed from small to large values, i.e., where
massive flavors smoothly decouple. On the other hand, this property
does not seem to be guaranteed in non-supersymmetric models,
though it is intuitively appealing and may well be quite generic.

Second, in more complicated models there may be fermions in the
fundamental theory that interact with background fields ${\bf A}^L(x)$
and remain massless even after the masses $m_0$ are introduced to
some of the flavors. In that case the adiabatic process discussed
above does not necessarily end up in the ground state (e.g.,
because  some energy levels of massless fermions cross zero). The precise
nature of the final state becomes a matter of complicated dynamics,
so the induced charge matching need not necessarily occur.

 The author is indebted to E.~Akhmedov, D.~Amati,
S.~Dubovsky, G. Dvali,
D.~Gorbunov, A.~Kuznetsov, M.~Libanov,
V.~Kuzmin, K.~Selivanov, A.~Smirnov,
P.~Tinyakov and S.~Troitsky for numerous
helpful discussions. This research was supported
in part under 
 Russian Foundation for Basic Research grant 96-02-17449a and by the
U.S.~Civilian Research and Development Foundation for
Independent States of FSU (CRDF) award RP1-187.
The author would like to thank Professor Miguel Virasoro
for hopsitality at the Abdus Salam International Center for Theoretical
Physics, where part of this work was carried out.

\end{document}